# Coherent manipulation of Andreev states in superconducting atomic contacts

C. Janvier[1], L. Tosi[1,2], L. Bretheau[1†], Ç. Ö. Girit[1††], M. Stern[1], P. Bertet[1], P. Joyez[1], D. Vion[1], D. Esteve[1], M. F. Goffman[1], H. Pothier[1], C. Urbina[1*].

[1] Quantronics Group, Service de Physique de l'État Condensé (CNRS UMR 3680), IRAMIS, CEA-Saclay, 91191 Gif-sur-Yvette, France

[2] Centro Atómico Bariloche and Instituto Balseiro, Comisión Nacional de Energía Atómica (CNEA), 8400 Bariloche, Argentina

[†]Present address: Department of Physics, Massachusetts Institute of Technology, Cambridge, Massachusetts 02215, USA

[††]Present address: CNRS USR 3573, Collège de France, 11 place Marcelin Berthelot, 75005 Paris, France

*Correspondence to: cristian.urbina@cea.fr

**Abstract**:

Coherent control of quantum states has been demonstrated in a variety of superconducting devices. In all these devices, the variables that are manipulated are collective electromagnetic degrees of freedom: charge, superconducting phase, or flux. Here, we demonstrate the coherent manipulation of a quantum system based on Andreev bound states, which are microscopic quasiparticle states inherent to superconducting weak links. Using a circuit quantum electrodynamics setup we perform single-shot readout of this "Andreev qubit". We determine its excited state lifetime and coherence time to be in the microsecond range. Quantum jumps and parity switchings are observed in continuous measurements. In addition to possible quantum information applications, such Andreev qubits are a testbed for the physics of single elementary excitations in superconductors.

The ground state of a uniform superconductor is a many-body coherent state. Microscopic excitations of this superconducting condensate, which can be created for example by the absorption of photons of high enough energy, are delocalized and incoherent because they have energies in a continuum of states. Localized states arise in situations where the superconducting gap $\Delta$ or the superconducting phase undergo strong spatial variations: examples include Shiba states around magnetic impurities (1), Andreev states in vortices (2) or in weak links between two superconductors (3). Because they have discrete energies within the gap, Andreev states are expected to be amenable to coherent manipulation (4,5,6,7,8). In the simplest weak link, a single conduction channel shorter than the superconducting coherence length $\xi$, there are only two Andreev levels $\pm E_A(\tau, \delta) = \pm \Delta \sqrt{1 - \tau \sin^2(\delta/2)}$, governed by the transmission probability $\tau$ of electrons through the channel and the phase difference $\delta$ between the two superconducting condensates (3). Despite the absence of actual barriers, quasiparticles (bogoliubons) occupying these Andreev levels are localized over a distance $\sim \xi$ around the weak link by the gradient of the superconducting phase, and the system can be considered an "Andreev quantum dot" (5,6). Figure 1 shows the energies $E_i(\delta)$ of the different states of this dot. In the spin-singlet ground state $|g\rangle$ only the negative energy Andreev level is occupied and $E_g = -E_A$. If a single quasiparticle is added, the dot reaches a spin-degenerate odd-parity state $|o\rangle$ with $E_o = 0$ (9-12). Adding a second quasiparticle of opposite spin to the dot in state $|o\rangle$ brings it to a spin-singlet even-parity excited state $|e\rangle$ with $E_e = +E_A$ (13,14). The $|e\rangle$ state can also be reached directly from $|g\rangle$ by absorption of a photon of energy $2E_A$. Here we demonstrate experimentally the manipulation of coherent superpositions of states $|g\rangle$ and $|e\rangle$, even if parasitic transitions to $|o\rangle$ are also observed.

Atomic-size contacts are suitable systems to address the Andreev physics because they accommodate a small number of short conduction channels (*15*). We create them using the microfabricated break-junction technique *(16)*. Figure 2 presents the sample used in the experiment. An aluminum loop with a narrow suspended constriction (Fig. 2C) is fabricated on a polyimide flexible substrate mounted on a bending mechanism cooled down to $\sim 30\,\mathrm{mK}$ *(17)*. The substrate is first bent until the bridge breaks. Subsequent fine-tuning of the bending allows creating different atomic contacts and adjusting the transmission probability of their channels. The magnetic flux $\phi$ threading the loop controls the phase drop $\delta = 2\pi\phi/\phi_0$ across the contact and thereby the Andreev transition frequency $f_A(\tau,\delta) = 2E_A/h$ ($\phi_0$ is the flux quantum, $h$ Plank's constant). To excite and probe the Andreev dot, the loop is inductively coupled to a niobium quarter-wavelength microwave resonator *(17)* (Fig. 2B) in a circuit quantum electrodynamics architecture *(18,19)*. The resonator is probed by reflectometry at frequency $f_0$ close to its bare resonance frequency $f_R \simeq 10.134\,\mathrm{GHz}$. The actual resonator frequency is different for each one of the three Andreev dot states: in the odd state, the resonance frequency is unaltered while the two even states lead to opposite shifts around $f_R$ *(20)*. The Andreev transition $|g\rangle \rightarrow |e\rangle$ is driven using a second tone of frequency $f_1$. Details of the setup are shown in figures S1 and S2 *(20)*.

Here we present data obtained on a representative atomic contact containing only one high transmission channel. Data from other contacts is shown in figures S6-S8. First, a two-tone spectroscopy is performed by applying a 13 µs driving pulse of variable frequency, immediately followed by a 1 µs-long measurement pulse $\left(f_0 \simeq 10.1337\,\mathrm{GHz}\right)$ probing the resonator with an amplitude corresponding to an average number of photons $\bar{n} \simeq 30$ (see Fig. 3A). Apart from the signal at $f_1 \simeq f_0$, the spectrum displays a resonance corresponding to the Andreev transition. The spectrum is periodic in flux, with period $\phi_0$, which allows calibrating the value of $\delta$ across the contact (Fig. S3). Fits of the measured lines for different contacts with the analytical form of $f_A(\tau, \delta)$ provide the transmission probability $\tau$ of highly transmitting channels with up to five significant digits, as well as the superconducting gap $\Delta/h = 44.3$ GHz of the aluminum electrodes.

The coupling between the resonator and the Andreev dot is evident from the avoided crossing between the two modes observed in single-tone continuous-wave spectroscopy (Fig. 3B). Fitting the data with the predictions of a Jaynes-Cummings model *(19,20)*, yields the coupling strength $g/2\pi = 74$ MHz at the two degeneracy points where $f_A = f_R$. Remarkably, the resonance of the bare resonator is also visible for all values of the phase, signaling that on the time scale of the measurement the Andreev dot is frequently in the odd state $|o\rangle$ *(10,12,21)*.

Figure 3C shows the histograms of the reflected signal quadratures *I,Q* for a sequence of 8000 measurement pulses taken at $\delta = \pi$, without and with a $\pi$ driving pulse. The results gather in three separate clouds of points demonstrating that a single measurement pulse allows discriminating the dot state. The normalized number of points in each cloud is a direct measurement of the populations of the three states. The two panels of Fig 3C show the population transfer between the two even states induced by the driving pulse. Continuous measurement of the state of the Andreev dot in absence of drive, reveals the quantum jumps *(22)* between the two even states and the changes of parity corresponding to the trapping and untrapping of quasiparticles in the dot (Fig. 3D). The analysis *(23)* of this real-time trace yields a parity switching rate of $\sim 50$ kHz *(20)*.

The coherent manipulation at $\delta = \pi$ of the two-level system formed by $|g\rangle$ and $|e\rangle$ is illustrated in Fig. 4. Figure 4A shows the Rabi oscillations between $|g\rangle$ and $|e\rangle$ obtained by varying the duration of a driving pulse at frequency $f_1 = f_A(\tau, \pi)$ (Movie S1). Figure 4B shows how the populations of $|g\rangle$ and $|e\rangle$ change when the driving pulse frequency $f_1$ is swept across the Andreev frequency $f_A(\tau, \pi)$. After a $\pi$-pulse the populations relax exponentially back to equilibrium with a relaxation time $T_1(\delta = \pi) \simeq 4\,\mu s$ (Fig. 4D). The Gaussian decay by 1/e of detuned Ramsey fringes (Fig. 4F) provides a measurement of the coherence time $T_2^*(\delta = \pi) \simeq 38\,\mathrm{ns}$. This short coherence time is mainly due to low-frequency (<MHz) fluctuations of the Andreev energy $E_A(\tau, \delta)$, as shown by the much longer decay time $T_2(\delta = \pi) \simeq 565\,\mathrm{ns} \gg T_2^*$ of a Hahn echo (Fig. 4G). Measurements at $\delta = \pi$ on other contacts with the same sample, with transmissions corresponding to a minimal Andreev frequency $3\,\mathrm{GHz} < f_A(\tau, \pi) < 8\,\mathrm{GHz}$, give $T_1$ mostly around 4 μs (up to 8.5 μs), $T_2^*$ around 40 ns (up to 180 ns) and $T_2$ around 1 μs (up to 1.8 μs), but no clear dependence of the characteristic times on $\tau$ is observed (Fig. S7 and S8).

Figure 4E shows the measured relaxation rate $\Gamma_1 = 1/T_1$ as a function of the phase $\delta$. The expected Purcell relaxation rate arising from the dissipative impedance seen by the atomic contact (dotted line in Fig. 4E) matches the experimental results only close to the degeneracy points where $f_A = f_R$, but is about five times smaller at $\delta = \pi$. Based on existing models we estimate that relaxation rates due to quasiparticles *(24-28)* and to phonons *(7,8,21)* are negligible. Empirically, we fit the data at $\delta = \pi$ by considering an additional phase-independent relaxation mechanism, which remains to be identified.

The linewidth of the spectroscopy line, which is a measure of the decoherence rate, shows a minimum at $\delta = \pi$ (Fig. 4C). The Gaussian decay of the Ramsey oscillations points to *1/f* transmission fluctuations as the main source of decoherence at $\delta = \pi$, where the system is insensitive to first order to flux noise *(28)*. Fluctuations of $\tau$ can arise from vibrations in the mechanical setup and from motion of atoms close to the contact. Figure 4C also shows the linewidths calculated assuming *1/f* transmission noise and both white and *1/f* flux noise *(20)*. The amplitude of the *1/f* transmission noise, $2.5 \times 10^{-6}\,\mathrm{Hz^{-1/2}}$ at 1 Hz, was adjusted to fit the measurement at $\delta = \pi$. The amplitudes of the white and 1/f flux noise were then obtained from a best fit of the linewidth phase dependence. The extracted *1/f* noise amplitude ($5\,\mu\phi_0\,\mathrm{Hz^{-1/2}}$ at 1 Hz) is a typical value for superconducting devices and has a negligible effect to second order *(29)*. The source of the apparent white flux noise $\left(48\,\mathrm{n}\phi_0\,\mathrm{Hz^{-1/2}}\right)$ is not yet identified.

The Andreev quantum dot has been proposed as a new kind of superconducting qubit *(5,6)*, which differs markedly from existing ones *(30)*. In qubits based on charge, flux, or phase *(30)* the states encoding quantum information correspond to collective electromagnetic modes, while in Andreev qubits they correspond to microscopic degrees of freedom of the superconducting condensate. Our results are a proof of concept of this new type of qubit. Further work is needed to understand fully the sources of decoherence and to couple several qubits in multi-channel contacts *(5,8)*. The Andreev quantum dot, with its parity sensitivity, is also a powerful tool to investigate quasiparticle-related limitations on the performance of superconducting qubits *(28,31,32)* and detectors *(33)*. Furthermore, our experimental strategy could be used to explore hybrid superconducting devices in the regime where Andreev states evolve into Majorana states *(33,35,36)*.

**Acknowledgments:** We have benefited from technical assistance by P. Sénat and P.-F. Orfila and discussions with G. Catelani, V. Shumeiko and A. Levy Yeyati. We acknowledge financial support by contract ANR-12-BS04-0016-MASH. Ç.Ö.G. was supported by the People Programme (Marie Curie Actions) of the European Union's Seventh Framework Programme (FP7/2007-2013) under REA grant agreement no. PIIF-GA-2011-298415, and L.T. by ECOS-SUD (France-Argentina) grant N° A11E04 and a scholarship from CONICET (Argentina).

Authors' contributions: C.J., M.F.G., H.P. and C.U. took part in all aspects of this work. L.T., L.B., Ç.Ö.G. and D.E. participated in the design of the experiment. M.S. and D.V. helped in setting up the microwave setup. L.T. and D.V. participated in running the experiment. All authors participated in interpreting the data and writing the manuscript.

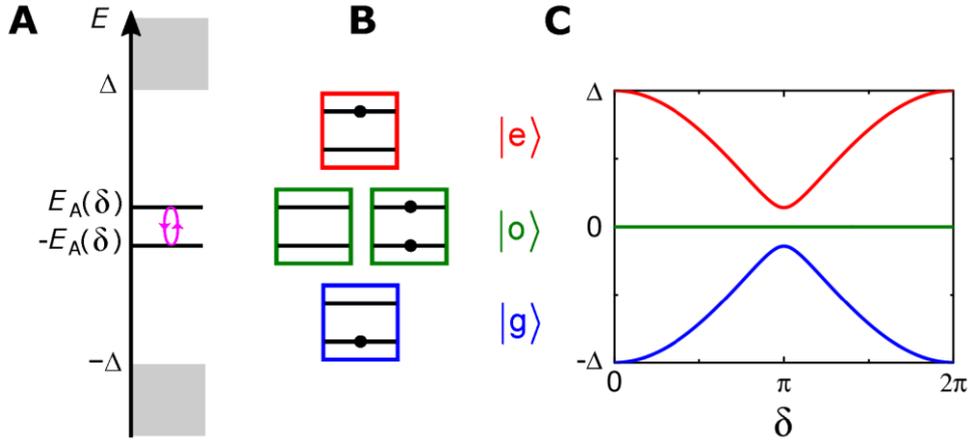

**Fig.1 Single channel Andreev quantum dot**. (**A**) Energy levels: Two discrete Andreev bound levels detach symmetrically from the upper and lower continua of states (light grey regions for $|E| > \Delta$). Photons of energy $2E_A$ can induce transitions between the two Andreev levels (magenta arrows). (**B**) Andreev levels occupation in the four possible quantum states of the Andreev dot. Only the lower Andreev level is occupied in the ground state $|g\rangle$ (blue box). In the excited state $|e\rangle$ (red box) only the upper Andreev bound level is occupied. In the doubly degenerate odd state $|o\rangle$ both Andreev levels are either occupied or empty. (**C**) Energy of the four Andreev dot states for a channel of transmission probability $\tau = 0.98$, as a function of the phase difference $\delta$ across the weak link.

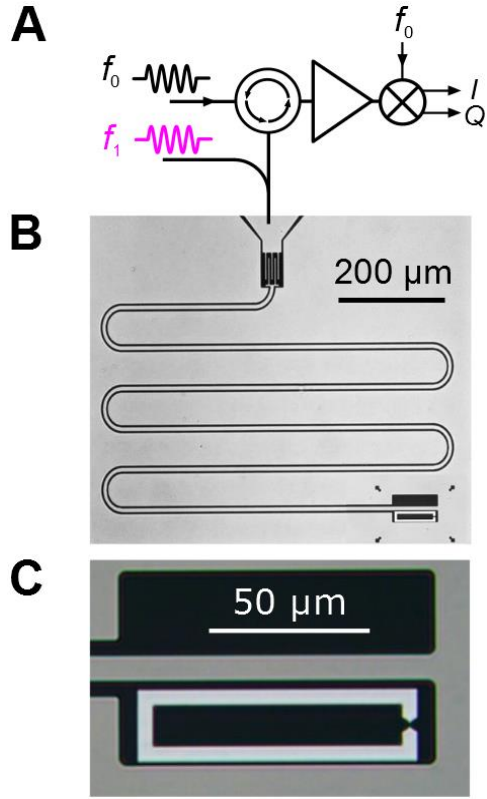

**Fig.2**. **Measurement setup of a superconducting atomic contact in a microwave resonator.** (**A**): Simplified 2-tone microwave setup. The measurement (frequency $f_0$) and drive (frequency $f_1$) signals are coupled to the resonator through the same port. After amplification the reflected signal at $f_0$ is homodyne detected by an IQ mixer and its two quadratures (I and Q) are digitized. (**B**): Optical micrograph of the $\lambda/4$ niobium coplanar meander resonator with an interdigitated capacitor $C \simeq 3\,\mathrm{fF}$ at the coupling port. At the shorted end an aluminum loop is inductively coupled to the resonator field. The resonator has resonance frequency $f_R \simeq 10.134\,\mathrm{GHz}$, with total quality factor $Q = 2200$, close to critical coupling (see Fig S4). (**C**): Detailed view of the aluminum loop. Upon bending the substrate the loop breaks at the narrow constriction to create an atomic contact.

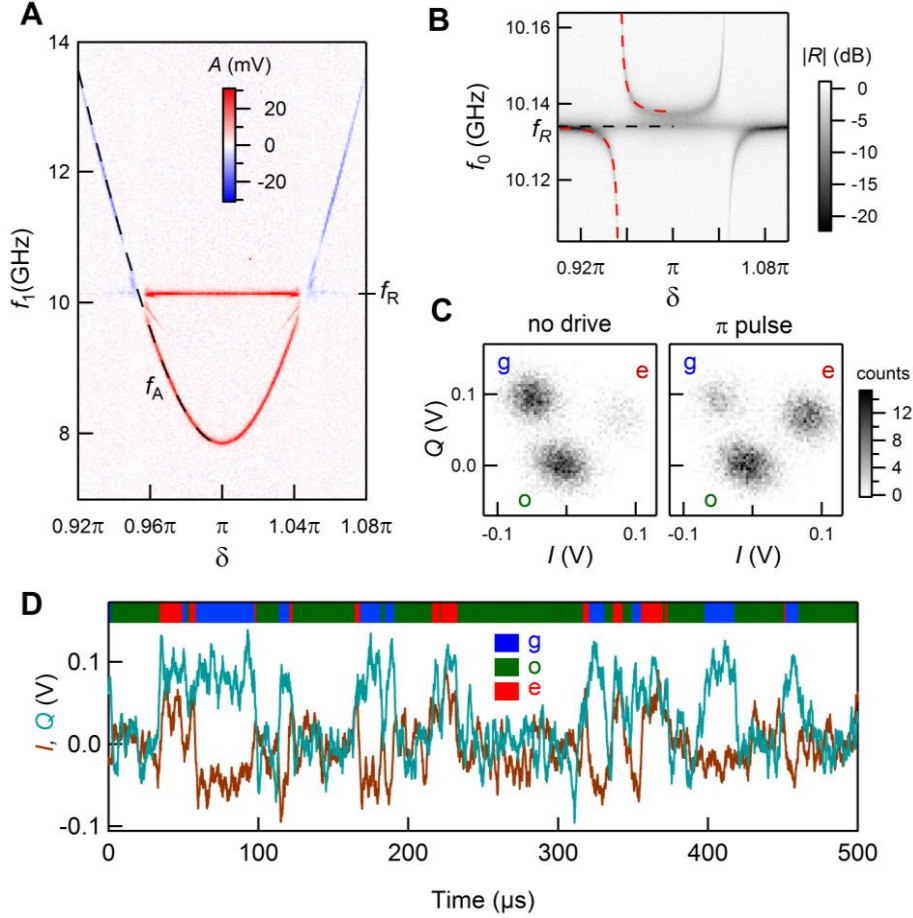

**Fig. 3. Spectroscopy and quantum jumps.** (**A**) Pulsed two-tone spectroscopy: color coded amplitude $A$ of one quadrature of reflected signal as a function of $\delta$ and $f_1$. Dashed black line: theoretical fit of Andreev transition frequency $f_A = 2E_A/h$ with $\tau = 0.99217$. A parasitic line, corresponding to a two photon process ($2f_1 = f_R + f_A(\tau, \delta)$), is visible just below $10\,\mathrm{GHz}$. (**B**) Single-tone continuous-wave spectroscopy using a vector network analyzer (average number of photons in resonator $\bar{n} \simeq 0.1$): resonator reflection amplitude $|R|$ as a function of $\delta$ and $f_0$. Red dashed curves: fits of the anti-crossings *(20)*. Data aligned with black dashed line correspond to the Andreev dot in state $|o\rangle$. (**C**) Histograms of $I, Q$ quadratures at $\delta = \pi$ illustrate single-shot resolution of the quantum state of the dot. Left panel: no drive at $f_1$. Right panel: $\pi$–pulse transfering population from $|g\rangle$ to $|e\rangle$. (**D**)

Continuous measurement at $\delta = \pi$, with $\bar{n} \simeq 100$ and no driving signal. Brown (cyan) time trace corresponds to $I$ ($Q$) quadrature. The color (blue, green, red) of the horizontal bar represents an estimate of the state (g, o, e) found using a hidden Markov Model toolbox (23).

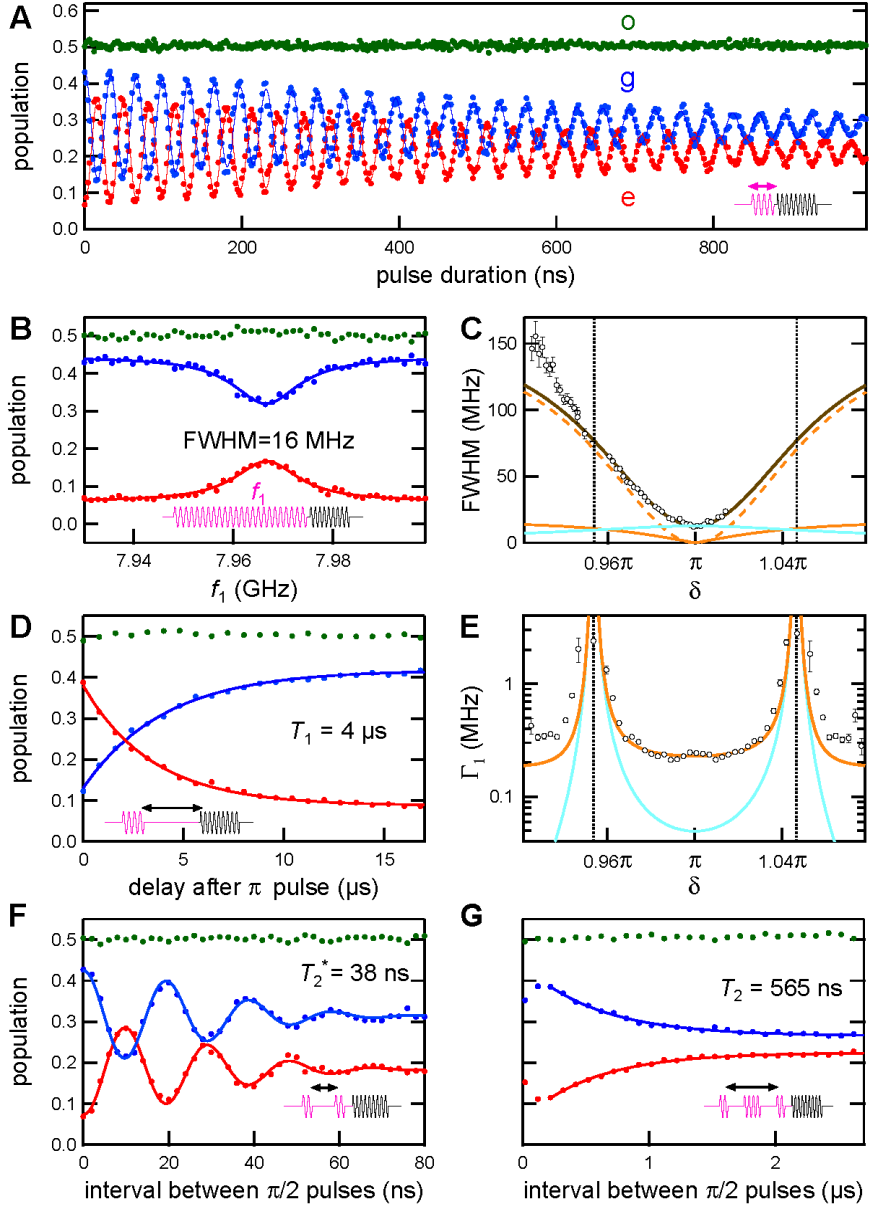

**Fig. 4. Coherent manipulation of Andreev quantum dot states at $\delta = \pi$.** Color dots show measured populations: ground (blue), excited (red) and odd (green) states. Lines are theoretical fits. Sketches of pulse sequences for each type of measurement are shown in each panel (magenta: drive; black: measurement). (**A**) Rabi oscillations: populations as a function

of the driving pulse duration. (**B**) Spectroscopy: populations as a function of driving pulse frequency $f_1$. (**C**) Phase dependence of linewidth (FWHM) of the spectral line. Dots: as extracted from a lorentzian fit of the experimental resonances (*20*). Black curve is best fit to the data, including the contributions of 1/f transmission noise (cyan line), and both 1/f (orange line) and white flux noise (orange dashed-line). Vertical dotted lines indicate phases for which $f_A = f_R$. (**D**) Relaxation of populations after a $\pi$ driving pulse. (**E**) Phase dependence of relaxation rate $\Gamma_1 = 1/T_1$. Dots: experimental data. Continuous curve is the sum of the expected Purcell rate (dotted line) and an empirical phase independent rate (180 kHz). (**F**) Ramsey fringes: populations as a function of delay between the two $\pi/2$-pulses detuned at $f_1 = f_A(\tau, \pi) + 51\,\mathrm{MHz}$. (**G**) Hahn echo: populations as a function of delay between the two $\pi/2$-pulses with a $\pi$-pulse in between.

# Supplementary Materials

## Materials and Methods

### Theoretical description of the system

The Hamiltonian of the system can be written as $H = H_A + H_R + H_{AR}$ , where the first term, the Andreev Hamiltonian, describes the atomic contact; the second one describes the electromagnetic resonator; and the third one accounts for the coupling between them. The Andreev Hamiltonian in the Andreev basis (*5*) is given by

$$H_A(\delta) = -E_A(\delta)\hat{\sigma}_z$$

where $\hat{\sigma}_z$ is a Pauli matrix acting in the $|g\rangle, |e\rangle$ space. The electromagnetic resonator is treated as a discrete single-mode oscillator described by

$$H_R = \hbar\omega_R(a^\dagger a + 1/2)$$

where $a^\dagger(a)$ are the creation (annihilation) photon operators. The term describing the coupling between the atomic contact and the resonator (up to first order) is given by

$$H_{AR} = M \hat{I}_R \hat{I}_A(\delta)$$

where $M$ is the loop-resonator mutual inductance, $\hat{I}_R = \sqrt{\hbar\omega_0/2L}\left(a^\dagger + a\right)$ is the transmission line current operator at the position of the atomic contact loop and $\hat{I}_A(\delta)$ is the Andreev current operator

$$\hat{I}_A(\delta) = I_A(\delta,\tau)\left[\hat{\sigma}_z + \sqrt{1-\tau}\tan(\delta/2)\hat{\sigma}_x\right]$$

with $I_A(\delta,\tau) = -\varphi_0^{-1}\partial E_A(\delta,\tau)/\partial\delta$ . As a result, in the region close to the degeneracy $hf_R = 2E_A$, where the rotating-wave approximation holds, the coupling Hamiltonian can be reduced to a Jaynes-Cummings model (*3*)

$$H_{AR} = \hbar g(\delta)\left(a\hat{\sigma}^+ + a^\dagger\hat{\sigma}^-\right)$$

where $\sigma^+ = |e\rangle\langle g|$ and $\sigma^- = |g\rangle\langle e|$. The phase dependent coupling energy $\hbar g(\delta)$ is given by

$$\hbar g(\delta) = \sqrt{z}\frac{\Delta}{2}\frac{E_A(\pi)}{E_A(\delta)}\tau\sin^2(\delta/2)$$

with $z = (M/L)^2 \pi Z_0/R_Q$ a constant coupling parameter. Fitting the anti-crossing depicted in Fig. 3B we obtain $g(\pi)/2\pi = 95.6 \,\text{MHz}$ and $z = 1.9\,10^{-5}$ .

## Relaxation rate through the resonator (Purcell effect)

Following Desposito and Yeyati (*4*) the relaxation rate due to the coupling to the environment can be estimated by using the expression

$$\Gamma_1 = \frac{\pi}{2}\frac{\Delta}{\hbar}\frac{\text{Re}\left[Z_{env}(\omega = 2E_A(\delta)\right]}{R_Q}\frac{(1-\tau)\left(\tau\sin^2\left(\delta/2\right)\right)^2}{\left(1-\tau\sin^2\left(\delta/2\right)\right)^{3/2}}.$$

In the phase region were $T_1$ was measured, the real part of the impedance seen from the atomic contact can be approximated by

$$\text{Re}\left[Z_{env}\left(\omega\right)\right] = \frac{z\,R_Q}{\pi}\frac{Q}{1+\left(Q\dfrac{1-\left(\omega/\omega_0\right)^2}{\omega/\omega_0}\right)^2}$$

where $Q$ and $\omega_0$ are the total quality factor and the resonant frequency of the resonator far from the anti-crossing.

## Fit of the resonances

In Fig. 4C, we compare with theory the measured linewidth of the Andreev resonance as a function of the phase. The experimental data were fitted with Lorentzian functions appropriate for white noise. However, for *1/f* noise theory predicts Gaussian resonances. The combination of the contributions of the three considered noise sources (*1/f* transmission noise, white and *1/f* flux noise) leads to a lineshape which is a convolution of a Lorentzian and a Gaussian function. In order to compare with experiment, we proceeded as for the experimental data and fitted the calculated resonance with a Lorentzian function on a 300 MHz interval, to extract a linewidth.

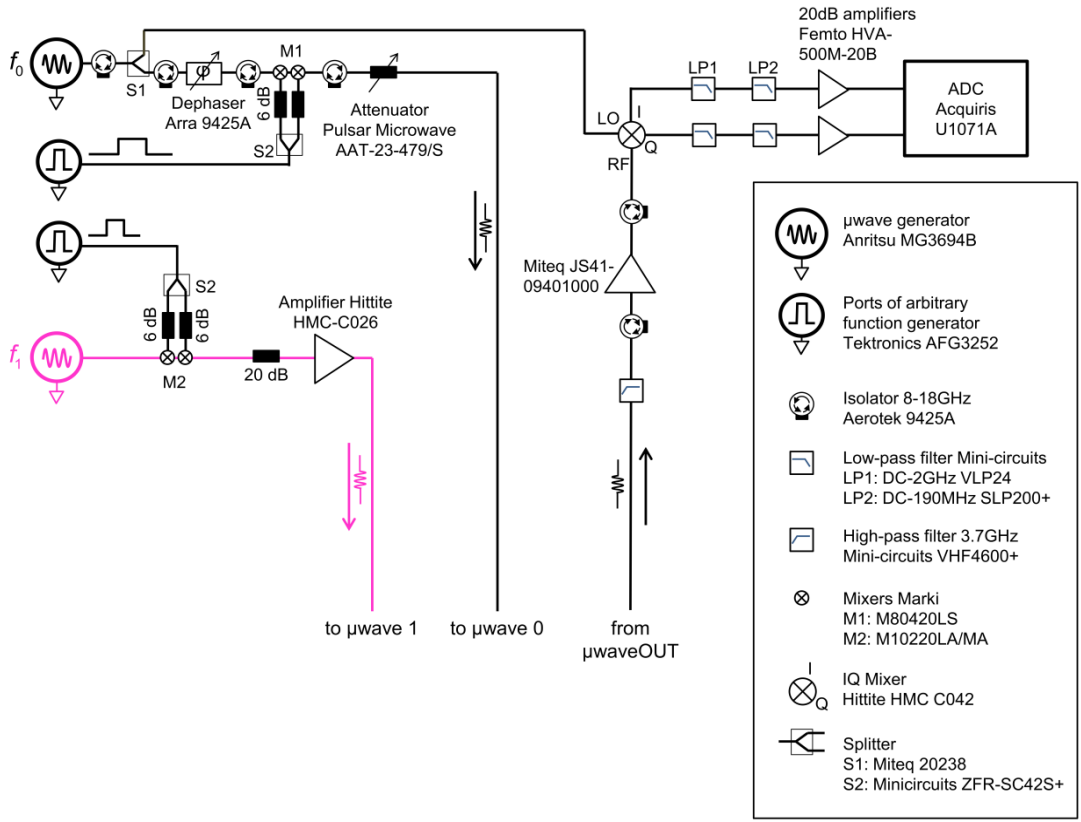

**Fig. S1:** Microwave setup at room temperature. There are two lines to inject driving ("µwave1") and measurement ("µwave0") pulses, and one line ("µwaveOUT") that carries the reflected signal at the measurement frequency. Microwave pulses are shaped by mixing continuous waves from the microwave sources with DC pulses from a 2-port arbitrary function generator. The latter and the acquisition board (ADC) are synchronized and triggered by an arbitrary waveform generator (Agilent AWG 33250, not represented). In order to improve the ON-OFF contrast of the microwave pulses, a second AWG (not represented) is used to pulse the $f_1$ microwave source itself.

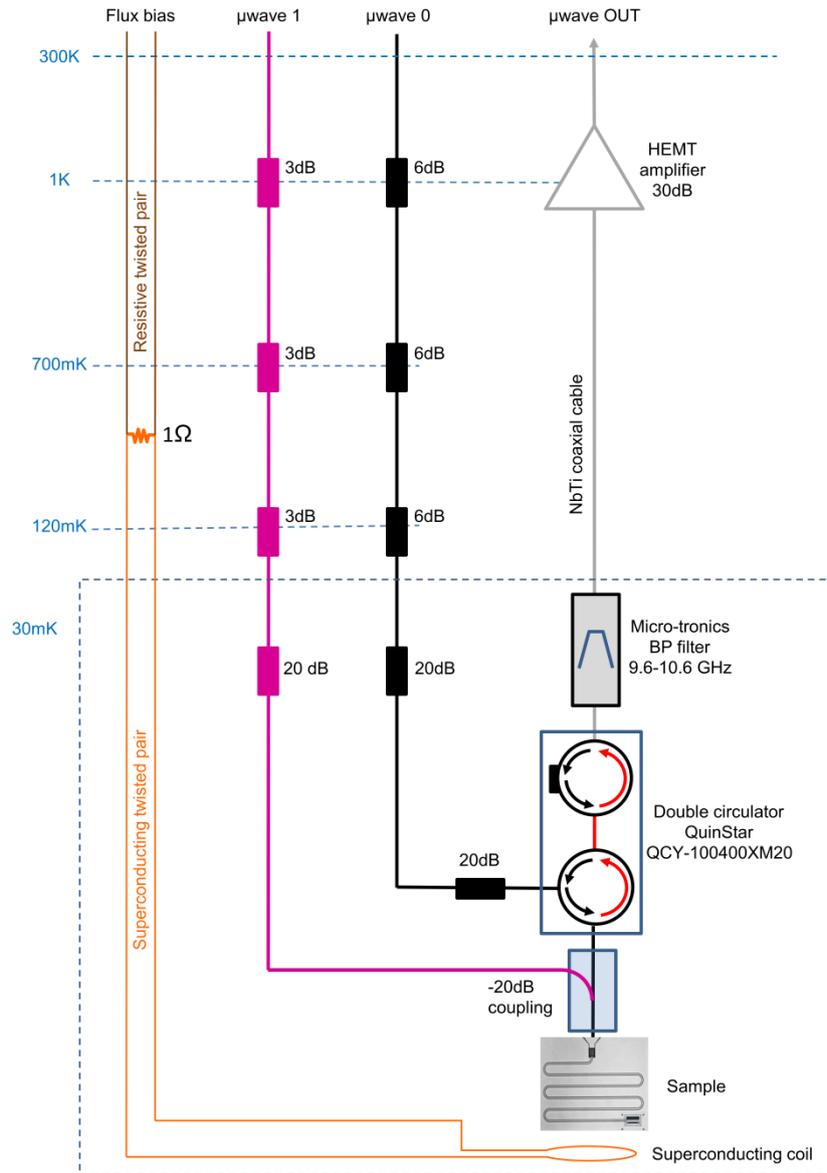

**Fig. S2:** Low temperature wiring. The three lines "µwave1", "µwave0" and "µwaveOUT" correspond to those of Fig. S1. The sample is enclosed in four shields: the inner one is made out of epoxy loaded with brass and carbon powder, the 2ⁿᵈ one out of aluminum, the 3ʳᵈ one out of Cryoperm, the 4ᵗʰ one out of copper. The sample and the shields are thermally anchored to the mixing chamber of base temperature 30 mK. The cryogenic microwave amplifier is a commercial HEMT (CITCRYO1-12A-1 from Caltech) with nominal gain 32 dB and noise temperature 7 K at 10 GHz. A DC magnetic field is applied perpendicular to the chip using a small superconducting coil placed a few mm above the aluminum loop containing the atomic contact. Biasing is performed using a voltage source (iTest BILT BE2102) in series with a 200 kΩ resistor. Filtering is provided partly by a 1 Ω resistor placed at 0.7 K in parallel with the coil.

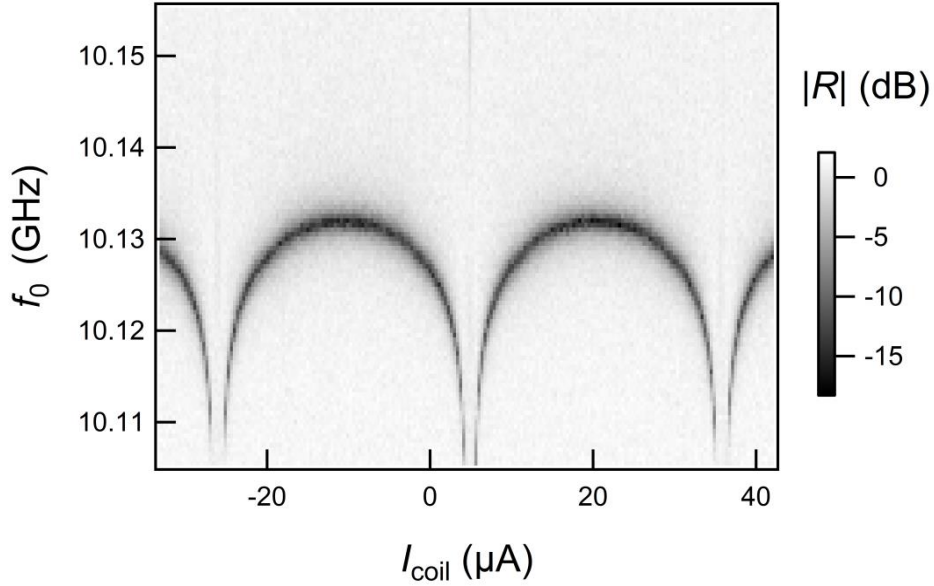

**Fig. S3: Periodicity of VNA measurements with flux.** Modulus $|R|$ of reflected signal as a function of the current $|I_{coil}|$ through the superconducting coil for a contact with several channels. The period allows calibrating the current associated with one flux quantum in the aluminum loop, i.e. with a $2\pi$ change in the phase $\delta$ across the contact. The currents at which the resonance frequency (dark) presents broad maxima correspond to $\delta = 0$ modulo $2\pi$.

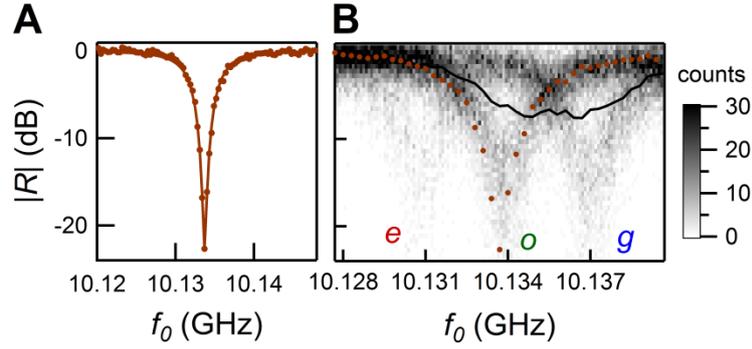

**Fig. S4: Vector network analyser (VNA) measurements** of the resonator for the contact with $\tau = 0.99217$ described in the manuscript. **(A)** Amplitude $|R|$ of reflexion coefficient as a function of the probe frequency $f_0$ when the resonator and the Andreev doublet are far detuned ($\delta = 0.9\pi$, $f_A \simeq 15.9\,\mathrm{GHz}$), corresponding to a vertical cut on the left edge of Fig. 3B. Symbols: measurement acquired at low power, corresponding to $\bar{n} \simeq 0.1$ photons in the resonator and a $10\,\mathrm{Hz}$ acquisition bandwidth. Solid line: fit using $|\mathrm{R}|^2 = 1 - \dfrac{1 - q^{-1}}{Qx^2 + q/4}$ with $x = f_0/f_R - 1$ and $q = Q_{ext}/Q$. The dip signals the resonance frequency $f_R \simeq 10.134\,\mathrm{GHz}$, with total quality factor $Q = 2200$ and external quality factor $Q_{ext} = 4800$ associated with the coupling capacitor. **(B)** $|R|$ measured at $\delta = \pi$. Black curve is taken at low power ($\bar{n} \simeq 0.1$ photons in resonator) and a $10\,\mathrm{Hz}$ acquisition bandwidth (corresponds to a cut in the middle of Fig. 3B). Image in the background is a two-dimensional histogram of 32000 data points taken in a single frequency sweep with a $600\,\mathrm{kHz}$ bandwidth, and a larger power ($\bar{n} \simeq 40$ at resonance). We observe three replicas of the resonance as measured at $\delta = 0.9\pi$ (brown symbols, same data as (A)). The central one corresponds to odd state $|o\rangle$, the rightmost to $|g\rangle$ and the leftmost (barely visible) to $|e\rangle$.

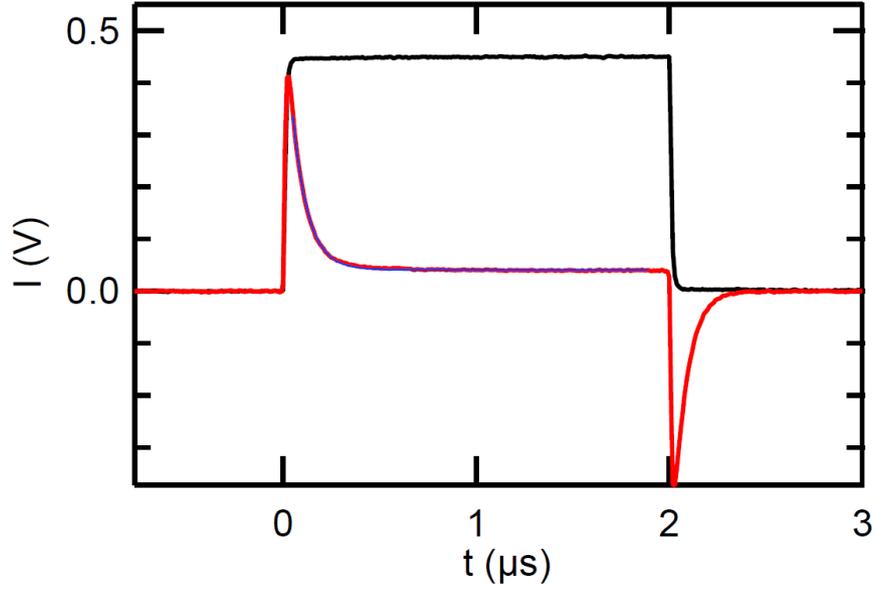

**Fig. S5:** Time-resolved response of the resonator to a 2 μs-long probe pulse. Black: 0.1 GHz-detuned pulse: complete reflection. Red: pulse at resonance frequency. After a loading time $2/\kappa$ of the cavity, wave exiting the cavity interferes destructively with reflected wave. The negative signal after t=2μs corresponds to photons exiting the cavity after the end of the pulse. Blue: exponential fit, with decay time $2/\kappa = 69\,\text{ns}$. Total quality factor of the cavity is $Q = \omega/\kappa = 2200$ in agreement with fit of cavity resonance (see Fig. S4**A**).

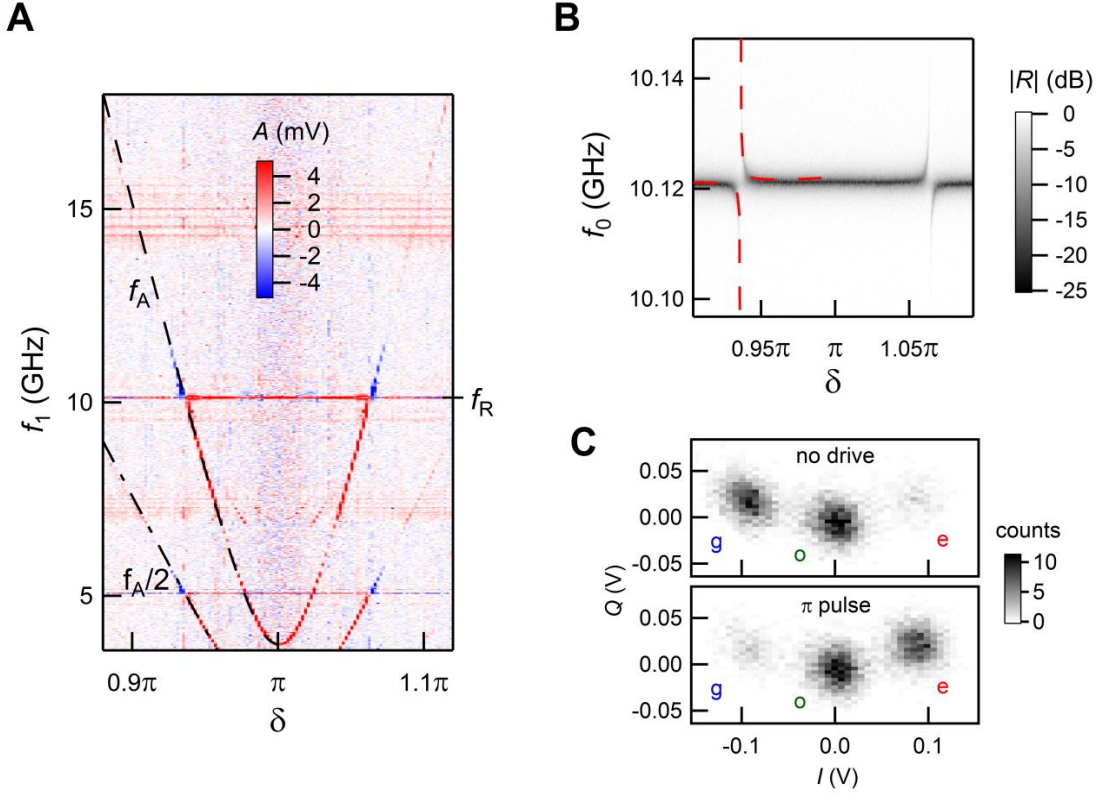

**Fig. S6:** Data for an atomic contact different from the one in the main text. (**A**) Pulsed two-tone spectroscopy: color coded amplitude $A$ of one quadrature of reflected signal as a function of $\delta$ and $f_1$. Dashed black line: theoretical fit of Andreev transition frequency $f_A = 2E_A/h$ with $\tau = 0.99806$. Two-photon processes (dash-dotted line labelled $f_A/2$) are observed because a higher excitation power than the one used for Fig. 3. (**B**) Single-tone continuous-wave spectroscopy using a vector network analyzer ($\bar{n} \simeq 0.1$): resonator reflection amplitude $|R|$ as a function of $\delta$ and $f_0$. Red dashed curves: fits of the anti-crossings using $g(\pi)/2\pi = 72\,\mathrm{MHz}$. Compared to Fig. 3, this data was taken on a different cool-down of the sample, and the bare resonator frequency was 10.121 GHz. (**C**) Density plots of $I$, $Q$ quadratures at $\delta = \pi$ illustrate single-shot resolution of the quantum state of the dot. Top panel: no drive at $f_1$. Bottom panel: $\pi$–pulse results in a population transfer from $|g\rangle$ to $|e\rangle$.

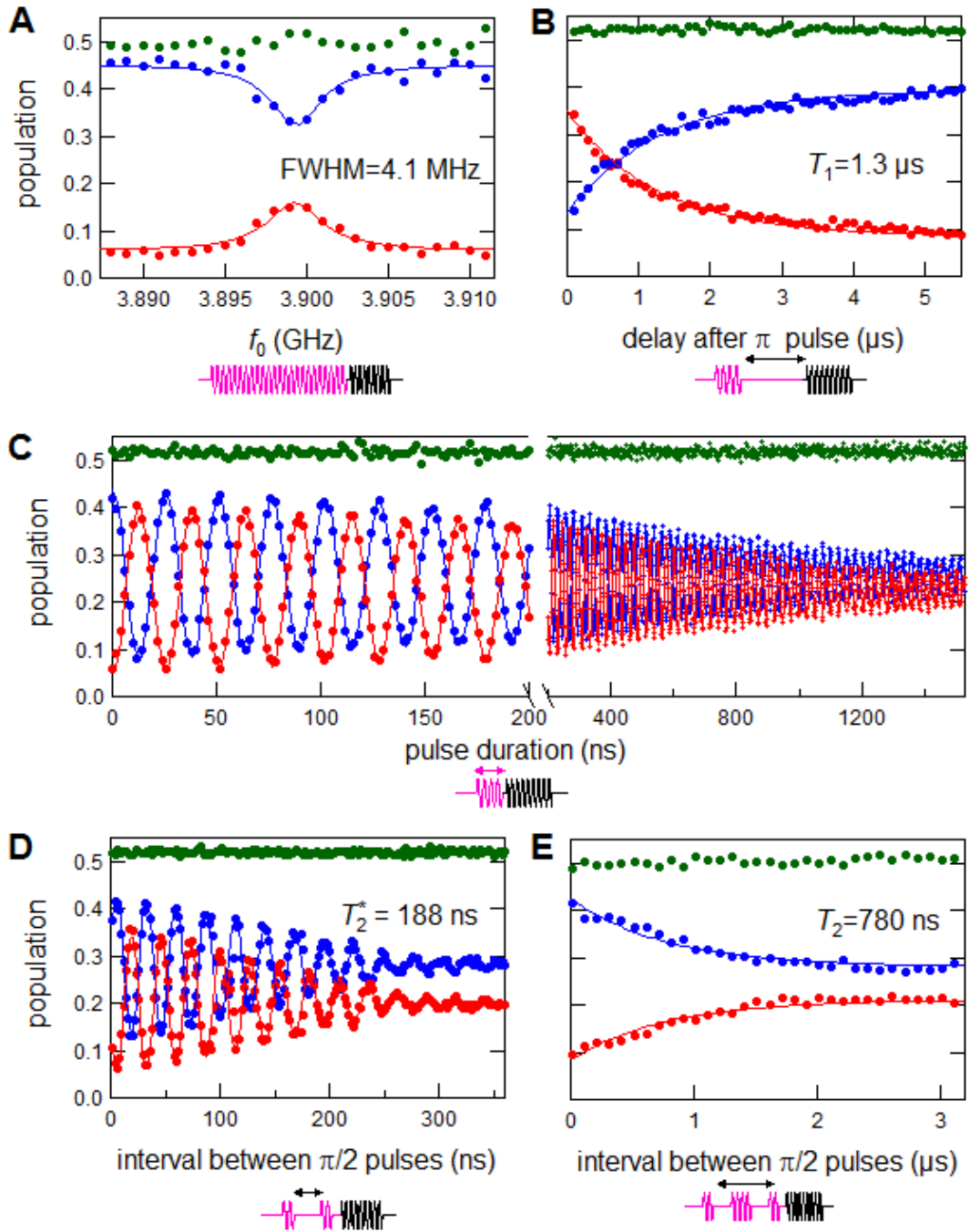

**Fig. S7:** Data for same contact as in Fig. S6, to be compared with Fig. 4. (**A**) Spectroscopy. (**B**) relaxation after a π-pulse. (**C**) Rabi oscillations (note break and change in scale of x-axis). (**D**) Ramsey fringes with 50 MHz detuning. (**E**) Hahn echo.

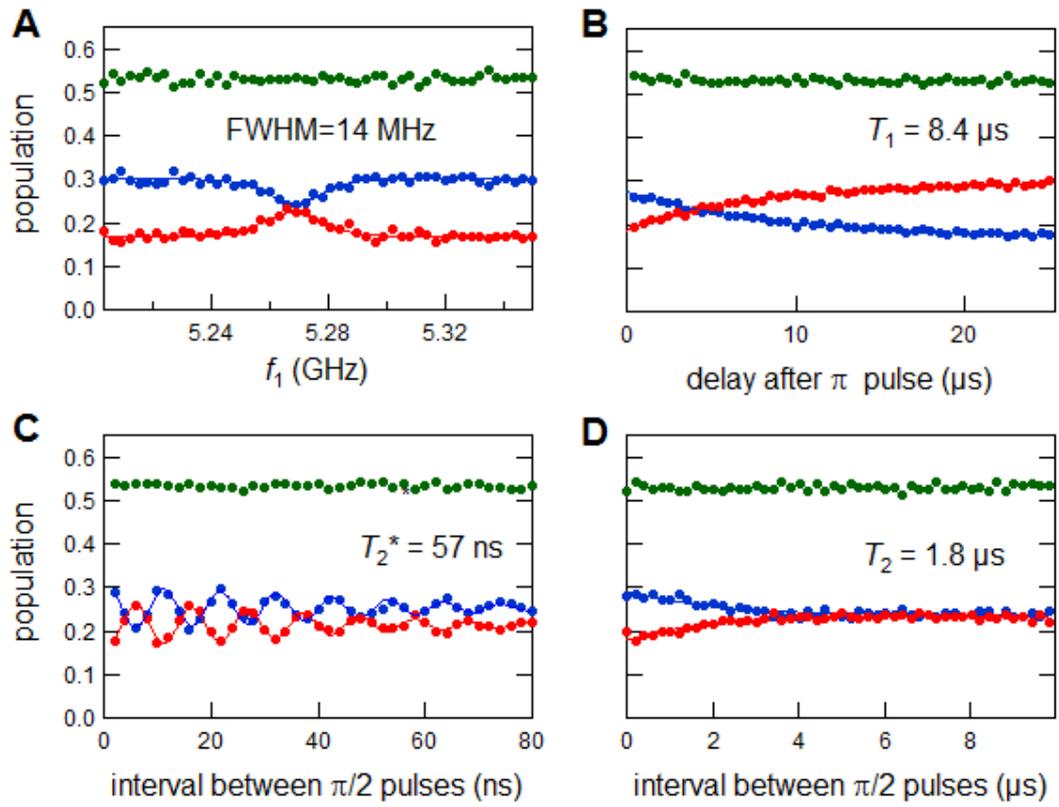

**Fig. S8:** Data for contact with channel transmission $\tau = 0.99647$, to be compared with Fig. 4. **(A)** Spectroscopy. **(B)** relaxation after a π-pulse. **(C)** Ramsey fringes with 95 MHz detuning. **(D)** Hahn echo.

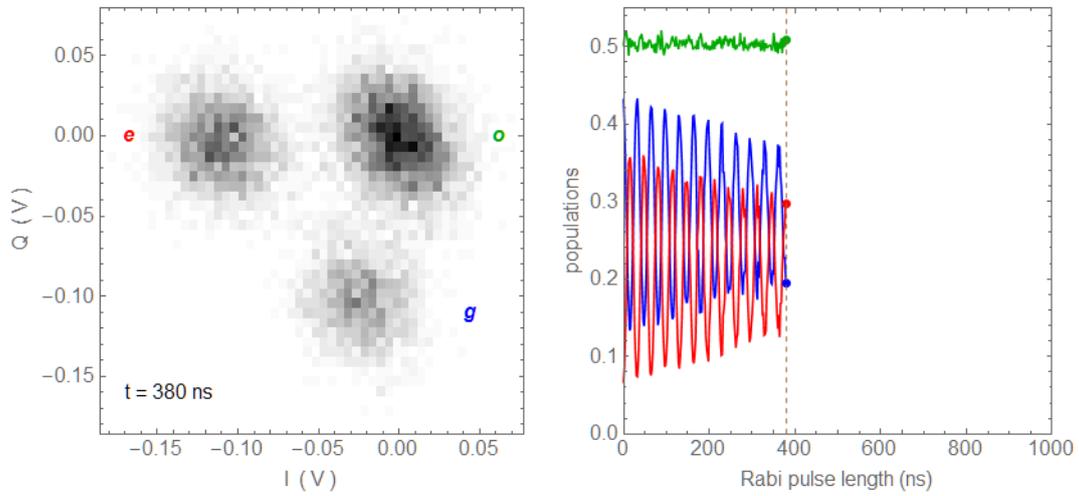

**Fig. S9 (snapshot from Movie S1):** (Left) Density plot of *I* and *Q* quadratures for a 380 ns-long Rabi pulse. (Right) Evolution of the populations of states $|g\rangle$ (blue), $|e\rangle$ (red) and $|o\rangle$ (green) for Rabi pulse lengths up to 380 ns.

**Movie S1:** (animated Gif image, available on the Science website) Rabi oscillations seen in the *I,Q* plane, and corresponding evolution of the populations: the populations of the ground state ($|g\rangle$) and the excited state ($|e\rangle$) swap, whereas the population of the odd state ($|o\rangle$) remains constant. Data correspond to Fig. 3C of paper, with a rotation in the *I,Q* plane.